\documentclass[
	aps,
	amsmath,amssymb,
	reprint,
	superscriptaddress
   ]{revtex4-1}

\usepackage{graphicx}
\usepackage{color}
\usepackage{dcolumn}
\usepackage{siunitx}
\usepackage{bm}
\usepackage{braket}

\begin{document}

\title[]{Calculating transition amplitudes by variational quantum deflation}

\author{Yohei Ibe}
\email{ibe@qunasys.com}
\affiliation{QunaSys Inc., Aqua Hakusan Building 9F, 1-13-7 Hakusan, Bunkyo, Tokyo 113-0001, Japan}

\author{Yuya O. Nakagawa}
\affiliation{QunaSys Inc., Aqua Hakusan Building 9F, 1-13-7 Hakusan, Bunkyo, Tokyo 113-0001, Japan}

\author{Nathan Earnest}
\affiliation{IBM Quantum, IBM T. J. Watson Research Center, Yorktown Heights, NY 10598
}

\author{Takahiro Yamamoto}
\affiliation{QunaSys Inc., Aqua Hakusan Building 9F, 1-13-7 Hakusan, Bunkyo, Tokyo 113-0001, Japan}

\author{Kosuke Mitarai}
\affiliation{Graduate School of Engineering Science, Osaka University, 1-3 Machikaneyama, Toyonaka, Osaka 560-8531, Japan}
\affiliation{QunaSys Inc., Aqua Hakusan Building 9F, 1-13-7 Hakusan, Bunkyo, Tokyo 113-0001, Japan}

\author{Qi Gao}
\affiliation{Mitsubishi Chemical Corporation, Science \& Innovation Center, 1000, Kamoshida-cho, Aoba-ku, Yokohama 227-8502, Japan}

\author{Takao Kobayashi}
\affiliation{Mitsubishi Chemical Corporation, Science \& Innovation Center, 1000, Kamoshida-cho, Aoba-ku, Yokohama 227-8502, Japan}

\date{\today}

\begin{abstract}
	Variational quantum eigensolver (VQE) is an appealing candidate for the application of near-term quantum computers.
A technique introduced in [Higgot \textit{et al.}, Quantum \textbf{3}, 156 (2019)], which is named variational quantum deflation (VQD), has extended the ability of the VQE framework for finding excited states of a Hamiltonian.
However, no method to evaluate transition amplitudes between the eigenstates found by the VQD without using any costly Hadamard-test-like circuit has been proposed despite its importance for computing properties of the system such as oscillator strengths of molecules.
Here we propose a method to evaluate transition amplitudes between the eigenstates obtained by the VQD avoiding any Hadamard-test-like circuit.
Our method relies only on the ability to estimate overlap between two states, so it does not restrict to the VQD eigenstates and applies for general situations.
To support the significance of our method, we provide a comprehensive comparison of three previously proposed methods to find excited states with numerical simulation of three molecules (lithium hydride, diazene, and azobenzene) in a noiseless situation and find that the VQD method exhibits the best performance among the three methods.
Finally, we demonstrate the validity of our method by calculating the oscillator strength of lithium hydride, comparing results from numerical simulations and real-hardware experiments on the cloud enabled quantum computer IBMQ Rome. Our results illustrate the superiority of the VQD to find excited states and widen its applicability to various quantum systems.

\end{abstract}

\pacs{Valid PACS appear here}
\maketitle

\section{Introduction}

We are now in an era where quantum computing practitioners can regularly use noisy quantum computers with tens of qubits \cite{Preskill2018}.
Although the existing quantum computers have no immediate practical applications, we believe that there is a possibility that such a device outperforms existing classical algorithms in specific tasks.
Various quantum algorithms for near-term devices have been suggested recently~\cite{Farhi2014, Mitarai2018, Farhi2018, Havlicek2019, Kusumoto2019, McArdle2018, Cao2019, xu2019variational, huang2019near, bravo2019variational}, and among such, the variational quantum eigensolver (VQE) \cite{Peruzzo2014} is considered to be an appealing candidate for applications of near-term quantum computers.

The original VQE \cite{Peruzzo2014} is a method for constructing an approximate ground state of a Hamiltonian on a programmable quantum device based on the variational principle of quantum mechanics.
The VQE constructs the approximate ground state by iteratively tuning a quantum circuit to minimize an energy expectation value of the generated state.
Because the near-term quantum devices are believed to be capable of generating a wavefunction that is not classically achievable, the VQE has the potential to explore a variational space that has not been investigated before.
To expand the potential application of the VQE other than for the ground state, a lot of works have extended the method to evaluate properties of excited states of a target Hamiltonian \cite{Nakanishi2019, Parrish2019, Higgott2019, McClean2017, Ollitrault2019, Jones2019}.
Those methods generally inherit the iterative and variational feature of the VQE, i.e., they also iteratively optimize a quantum circuit relative to some cost function.

The major and perhaps popular algorithms among such extensions are the subspace-search VQE (SSVQE) \cite{Nakanishi2019}, the multistate contracted VQE (MCVQE) \cite{Parrish2019}, and the variational quantum deflation (VQD) \cite{Higgott2019}.  There are pseudo-eigenvalue extensions of VQE such as quantum subspace expansion (QSE) \cite{colless2017robust} or quantum equation of motion (qEOM) \cite{ollitrault2020quantum}, but these algorithms are not fully variational methods and are therefore not included in our comparisons. While the SSVQE and the MCVQE can readily evaluate the transition amplitude $|\braket{\psi_1|A|\psi_2}|^2$ of an observable $A$ with respect to two approximate eigenstates $\ket{\psi_1}$ and $\ket{\psi_2}$ in a hardware-friendly manner (i.e., using less-costly quantum gates and circuits), the VQD, to the best of our knowledge, lacks such a method for evaluating the transition amplitude~\footnote{In this paper, we call $|\braket{\psi_1|A|\psi_2}|^2$ a transition amplitude, rather than $\braket{\psi_1|A|\psi_2}$}.
Since the transition amplitude is related to properties of the system such as the absorption and emission spectrum of photons~\cite{Sakurai2017, Turro2009}, this severely limits the application range of the VQD method.

In this work, we fill this gap by providing a technique to evaluate the transition amplitude without using costly quantum circuits such as the Hadamard test~\cite{MitaraiMethodology}.
Our technique is not only for the VQD, but can also be applied in a general setting where we have two orthogonal states $\ket{\psi_1}$ and $\ket{\psi_2}$ and a means to evaluate the overlap $|\braket{\psi_1|\psi_2}|^2$, and wish to evaluate $|\braket{\psi_1|A|\psi_2}|^2$.
To support the significance of the proposed technique, we present a comprehensive comparison of the SSVQE, the MCVQE, and the VQD by conducting \textit{noiseless} numerical simulations, where we use exact energy expectation values in the optimization routine of the parametrized circuit.
In this test, we use molecular Hamiltonians of  LiH and two azo compounds: diazene and azobenzene (AB).
We find that, under this setting, the VQD generally exhibits better performance than the other two, which validates the importance of our proposed technique. Finally, as a demonstration of the technique, we conduct a proof-of-principle calculation both on a noisy simulator, i.e., expectation values of observables are simulated with the realistic hardware noise, including shot noise and environmental noise such as T1/T2 and readout errors, and on IBM's real quantum hardware.

\section{Evaluation of transition amplitudes}\label{sec:eval}

First, we propose a method to evaluate the transition amplitude of Hermitian operators between two quantum states.
More concretely, we present how to evaluate $\left|\braket{\psi_1|A|\psi_2}\right|^2$ for a Hermitian operator $A$ and quantum states $\ket{\psi_1}$ and $\ket{\psi_2}$ such that $\braket{\psi_1|\psi_2}=0$ and
\begin{align}
    A=\sum_{i}a_i P_i,
\end{align}
where $P_i\in\{I,X,Y,Z\}^{\otimes n}$. $I,X,Y,$ and $Z$ are Pauli operators and $a_i\in\mathbb{R}$.
We assume that, for any given two states $\ket{\psi_1}$ and $\ket{\psi_2}$, we can evaluate the overlap $|\braket{\psi_1|\psi_2}|^2$.
This evaluation can be performed by, e.g., the so-called swap test \cite{PhysRevLett.87.167902}.
Note that the requirements on the hardware of quantum computers to evaluate the overlap can be relaxed when we know quantum circuits $U_1$ and $U_2$ that generate $\ket{\psi_1}$ and $\ket{\psi_2}$, that is, we can evaluate the overlap by $|\braket{\psi_1|\psi_2}|^2=|\braket{0|U_1^\dagger U_2|0}|^2$.
This is the case for calculating the transition amplitude for approximate eigenstates obtained by the VQD.

Let us consider unitary gates
\begin{equation}
    U_{ij, \pm} = \frac{1}{\sqrt{2}}(I\pm iP_i)\frac{1}{\sqrt{2}}(I \pm iP_j),
\end{equation}
which can be realized as a product of Pauli rotation gates $U_{ij, \pm} = e^{\pm i\frac{\pi}{4}P_i} e^{\pm i\frac{\pi}{4}P_j}$. 
We can show the following equality holds under the assumption of $\braket{\psi_1|\psi_2}=0$,
\begin{widetext}
\begin{align}
    \left|\braket{\psi_1|A|\psi_2}\right|^2 =&\sum_{i} a_i^2 |\braket{\psi_1|P_i|\psi_2}|^2 \nonumber \\
    &+\sum_{i<j} a_i a_j \left[2\left|\braket{\psi_1|U_{ij,+}|\psi_2}\right|^2 + 2\left|\braket{\psi_1|U_{ij,-}|\psi_2}\right|^2 - |\braket{\psi_1|P_i|\psi_2}|^2 - |\braket{\psi_1|P_j|\psi_2}|^2 - |\braket{\psi_1|P_iP_j|\psi_2}|^2\right],
\label{eq: main transition formula}
\end{align}
\end{widetext}
which can be employed to evaluate $\left|\braket{\psi_1|A|\psi_2}\right|^2$.
More concretely, we can measure each term in the right-hand side of Eq.~\eqref{eq: main transition formula} on a quantum device by regarding it as a overlap between two states because $P_i, P_j, U_{ij,\pm}$ are unitary and $P_i\ket{\psi}, P_j\ket{\psi},U_{ij,\pm}\ket{\psi}$ can be realized on the device. 
We then combine the results of measurement according to the equation.
Note that the assumption $\braket{\psi_1|\psi_2}=0$ should always be satisfied.

Equation~\eqref{eq: main transition formula} is one of the main results of this work.
It reduces the evaluation of the transition amplitudes to a sequence of measurements of overlaps of two states. 
On the other hand, if we allow more complicated circuits to be executed on a device, we can construct an ancilla based technique to evaluate $\braket{\psi_1|A|\psi_2}$ as shown in Appendix \ref{appsec:ancilla}.

\section{Comparison of algorithms for excited states}

In this section, we compare the accuracy and capability of three algorithms for obtaining excited states of a given Hamiltonian on near-term quantum computers, namely the SSVQE, the MCVQE, and the VQD, by noiseless numerical simulations.
By ``noiseless'', we mean all of the expectation values required in the algorithms are exactly computed.
This situation is ideal compared with the real near-term devices but still appropriate for discerning the capability of the algorithms.
This section aims to support the significance of our method proposed in the previous section by showing that the VQD gives the best performance among the three, and given a sufficiently low error rate and large enough number of sampled shots, one can reasonbly expect that these results will remain consistent when run on an actual quantum processor.

For the comparison, we use electronic Hamiltonians of LiH, diazene, and AB molecule.
LiH is considered and employed as a simple ``benchmark" molecule by a variety of studies on quantum computational chemistry~\cite{Kandala2017, PhysRevA.99.062304, PhysRevX.8.031022, Grimsley2019}.
Diazene and AB, on the other hand, are more relevant to applications of quantum chemistry to industry.
In particular, AB is one of the most representative organic molecules which show cis-trans photoisomerization in photochemistry. 
AB has been attracting significant interests from the viewpoints of its photophysics and photochemistry associated with its various applications of photo-functional materials, and its derivatives are widely used as important photo-functional dyes in the industry \cite{hunger2007industrial}. 
For photo-functional molecules such as AB and its derivatives, it is crucial to theoretically predict their photophysical properties such as absorption and emission spectra or emission quantum yields and to elucidate their photochemical reaction mechanisms. 
Although the elucidation of its photoisomerization mechanism has been made theoretically so far, it remains controversial whether it proceeds with rotation or inversion or others.
The simulations presented here do not only support the significance of the proposed method, but can also be viewed as a first step toward the real-world application of near-term quantum devices.

\subsection{Settings of numerical simulation}

\begin{figure}
    \centering
    \includegraphics[width=\linewidth]{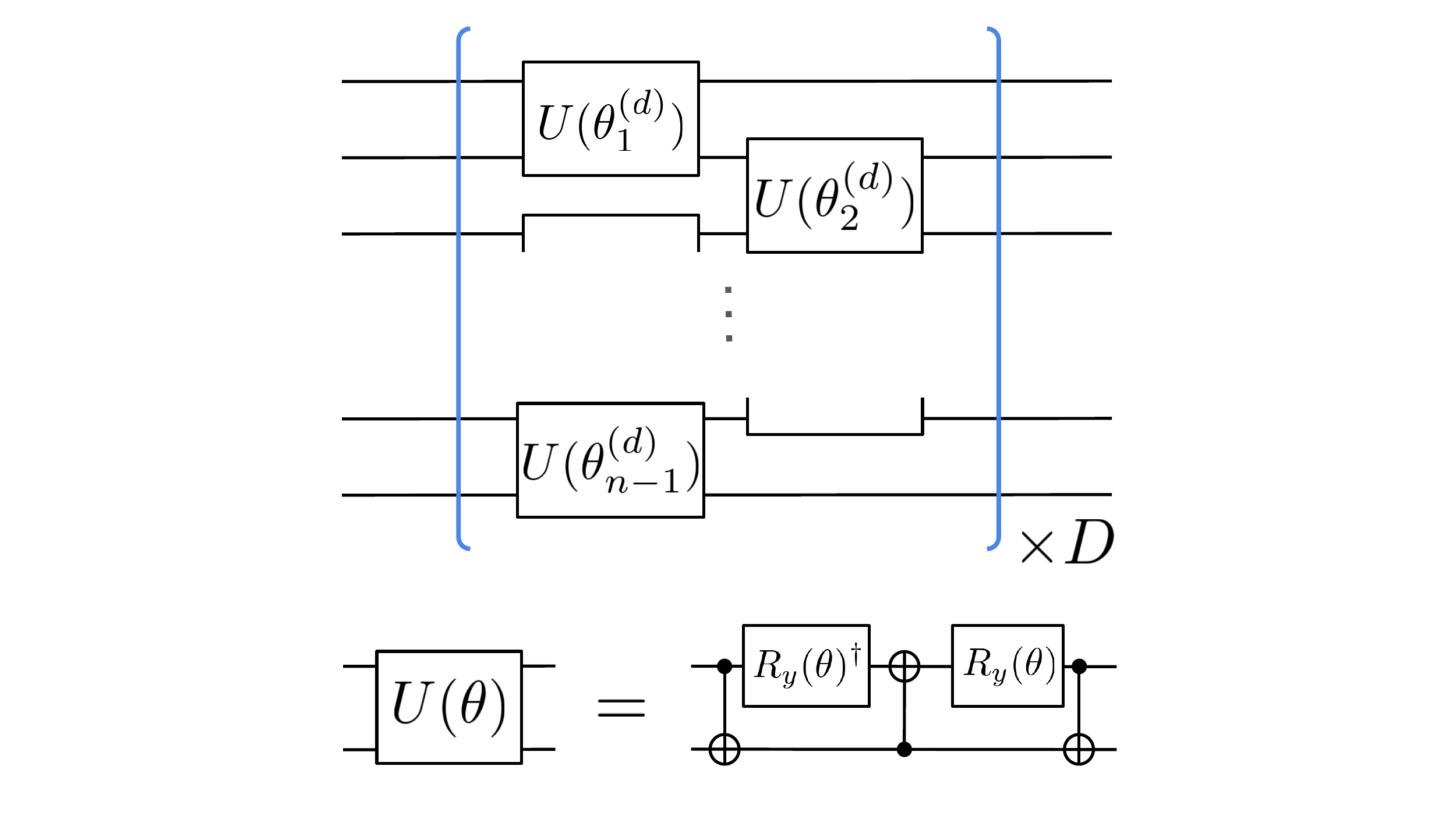} 
    \caption{Real-valued symmetry-preserving (RSP) ansatz. 
    In the figure, $R_y(\theta)=\exp(-i\theta Y/2)$. The rotation angles implemented in two-qubit unitary gates $U(\theta)$ are parameters to be optimized. $D$ denotes the depth of the circuit.}
    \label{fig:spr}
\end{figure}

Here we describe setups of numerical simulations for comparing the SSVQE, the MCVQE, and the VQD (the details of three algorithms are explained in Appendix~\ref{appsec:algorithms}).
As a variational quantum circuit for trial wavefunctions, we adopt an ansatz shown in Fig.~\ref{fig:spr} for all three algorithms.
We call this ansatz real-valued symmetry-preserving (RSP) ansatz, $U_{\mathrm{RSP}}(\bm{\theta})$, where $\bm{\theta}$ are classical parameters to be optimized.
It is a slightly modified version of a heuristic ansatz introduced in Ref. \cite{Barkoutsos2018}, which preserves the number of particles of a reference state, so that the generated wavefunction is always a real-valued vector in the computational basis.
As reference states for trial wavefunctions, we use the spin-restricted closed-shell Hartree-Fock state and singly excited states.
The BFGS method~\cite{Scipy2020} is employed for classical optimization of the ansatz quantum circuit.
The convergence criterion is set so that the optimization terminates when the relative difference of energy expectation value between iterations becomes lower than $10^{-8}$.
The electronic Hamiltonian is computed by PySCF \cite{PYSCF, PyscfRecent}, an open-source quantum chemistry library, and mapped to the qubit Hamiltonian by the Jordan-Wigner transformation \cite{Jordan1928}.
Note that the RSP ansatz only works for the Jordan-Wigner transformation.
For all simulations in this section, the weight vector for the SSVQE (see Appendix~\ref{appsec:ssvqe}) is set as $\bm{w} = (k, k-1, \ldots, 1)$, where $k$ is the number of quantum states to be calculated.

In the simulations, we compute several singlet and triplet eigenstates in the low-energy spectrum of the electronic Hamiltonians of LiH, diazene, and AB.
To circumvent the need to find all three degenerate eigenstates in each triplet subspace, we slightly modify the original electronic Hamiltonian $H$ to
\begin{equation}
    H' = H + \alpha S_z^2,
\end{equation}
where $S_z$ is an operator representing the $z$-component of the total electron spin, and $\alpha>0$.
When $\alpha$ is sufficiently large, all eigenstates of $H$ which have non-zero $S_z$ are projected out from the low-energy subspace of $H'$.
In the following subsections, we use this $H'$ with $\alpha=4$ in atomic units as the target Hamiltonian in the optimization process presented above.
This approach of adding ``penalty terms'' to a Hamiltonian can be found in Refs. \cite{McClean_2016, Ryabinkin2018, Kuroiwa2021}.
All simulations in this section are performed using Qulacs \cite{Qulacs}.

\subsection{Simple benchmark molecule: {\rm LiH}}

As for LiH molecule, we take STO-3G minimal basis set and all molecular orbitals into consideration, resulting in the number of simulated qubits being 12.
The RSP ansatz mentioned above with $D=10$ is used as the variational quantum circuit, where the total number of parameters is 110.
We calculate three energy levels S$_0$, T$_1$, and S$_1$~\footnote{In this paper, S$_i$ (T$_i$) ($i=0, 1,\cdots$) indicates the $i$-th singlet (triplet) eigenstate from the lowest energy eigenstate} for 36 points of the interatomic length of {\rm LiH} and compared with the full configuration interaction (full-CI) calculations.
As the initial values of ansatz parameters $\bm{\theta}$, we use uniform random numbers drawn from $[0, 2\pi]$ for the first point of the potential energy curve, and after that, we employ the optimized parameters of an adjacent point of the potential energy curve. 

\begin{figure}
  \begin{minipage}{1.0\linewidth}
    \centering
    \includegraphics[width=\linewidth]{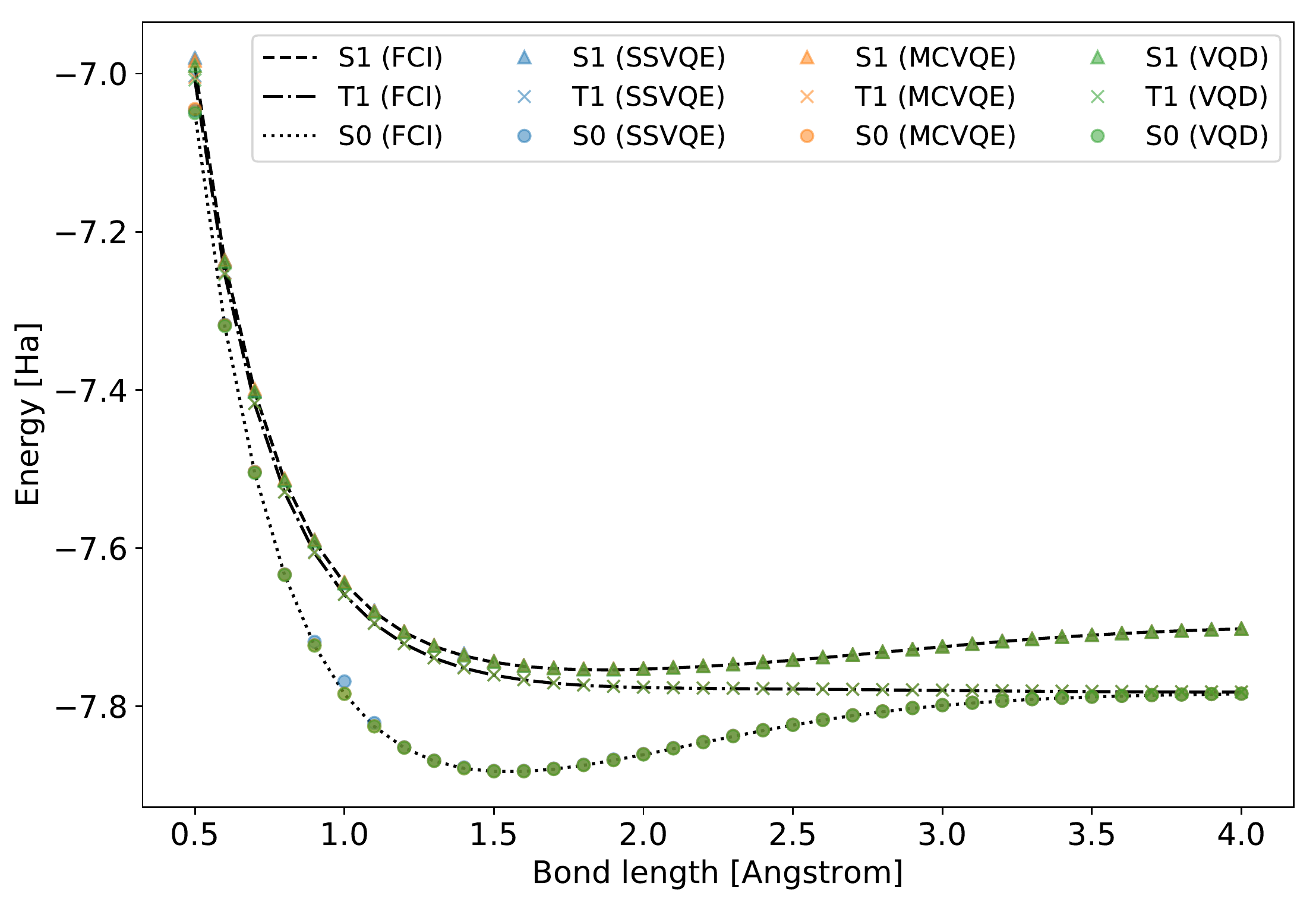}
  \end{minipage}\\
  \begin{minipage}{1.0\linewidth}
    \centering
    \includegraphics[width=\linewidth]{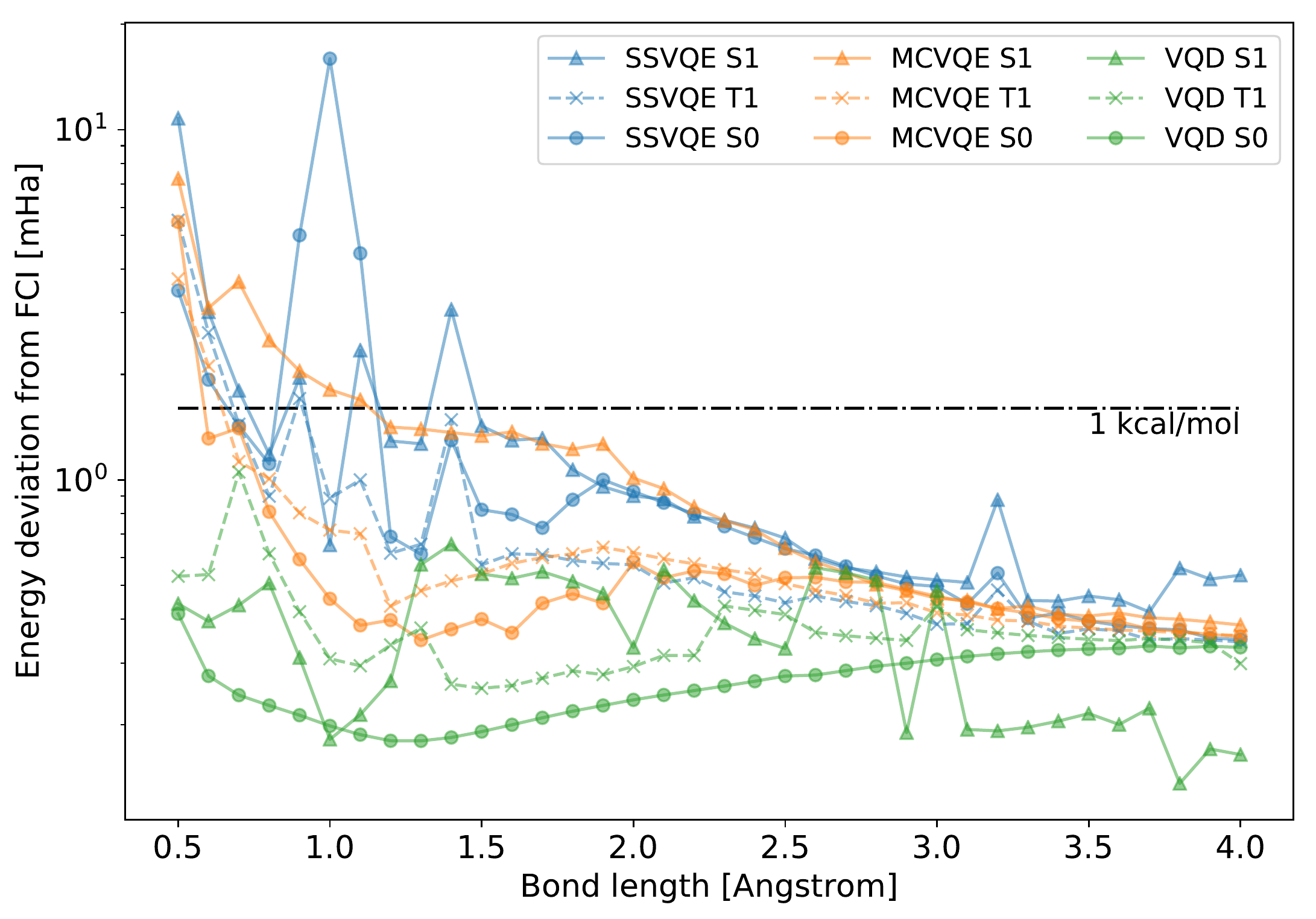}
  \end{minipage}
  \caption{Simulation of LiH molecule by the SSVQE, the MCVQE, the VQD, and the full configureation interactions (FCI). (top) The potential energy curve. Most of data points for three algorithms are overlapping.   (bottom) Energy deviations from the FCI calculations. The horizontal dashed line denotes the ``chemical accuracy".}\label{fig:lih_compare}
\end{figure}

Calculated energies and their errors of {\rm LiH} molecule by the SSVQE, the MCVQE, and the VQD are shown in Fig.~\ref{fig:lih_compare} along with the result of the full-CI calculations.
As seen in Fig. \ref{fig:lih_compare} (bottom), the VQD (the green lines) gives the most accurate energies than the other two, keeping the ``chemical accuracy" throughout the plot.

\subsection{More complex systems: diazene and azobenzene}

\subsubsection{Diazene}
For diazene, we present VQE simulations using molecular structures along its minimum energy path (MEP) between trans and cis isomers.
First, we obtain the structures by MEP calculations using the state-averaged complete active space self-consistent field (SA-CASSCF) method implemented in Molpro~2015~\cite{MOLPRO_brief} with the 6-31G* basis set.
The MEP calculations are done with the Gonzalez-Schlegel method~\cite{MEP_GonzalezSchlegel_1989}.
We use an active space consisting of 4 orbitals (2$\times$(lone pair on N) + $\pi$ (HOMO) + $\pi^*$ (LUMO)) with 6 electrons.
The S$_0$-S$_3$ and T$_1$-T$_3$ states are averaged in the SA-CASSCF calculations.
In the simulations, we use several points on S$_2$ MEP, from S$_2$ Franck-Condon (FC) state of trans/cis isomers to the S$_2$ minimum, which proceeds with a rotation about the N-N bond associating the disruption of the double bond.
Then, we perform the SSVQE, the MCVQE, and the VQD simulations of the Hamiltonians constructed in the same active space along the MEP and compare them with complete active space configuration interaction (CASCI) calculations, i.e., energies obtained by the exact diagonalization within the active space.
The number of qubits to be simulated is eight.
Here we utilize the RSP ansatz mentioned above with $D=20$, where the total number of parameters is 140.
Again, the initial values of ansatz parameters are taken as uniform random numbers drawn from $[0, 2\pi]$ for the first point on the MEP.
For other points on the MEP, optimized parameters at an adjacent point are used as initial values of parameters.

In addition to the energy of each eigenstate, we also calculate oscillator strength $f_{ij}$ between each pair of spin-singlet states $\ket{{\rm S}_i}$ and $\ket{{\rm S}_j}$.
It is defined as
\begin{equation}
    f_{ij} = \frac{2}{3}(E({\rm S}_j) - E({\rm S}_i)) \sum_{\alpha=x,y,z} |\braket{{\rm S}_j | R_\alpha | {\rm S}_i}|^2,
\label{eq: def os}
\end{equation}
where $E({\rm S})$ is the energy of $\ket{{\rm S}}$, $R_\alpha = \sum_{l=1}^N r_{l, \alpha}$ is the electric dipole moment operator, and $r_{l, \alpha}$ is the $\alpha$-coordinate of the $l$-th electron.
The oscillator strength gives the normalized strength of the absorption and emission spectrum of molecules~\cite{Turro2009}, so it is fundamental for studying photochemical dynamics and reactions of molecules in quantum chemistry.
Note that the oscillator strength involves the transition amplitude of $R_\alpha$ operator, so it is impossible to evaluate it by the VQD on a quantum device in a hardware-friendly manner without using our proposed method in Sec.~\ref{sec:eval}.
In this subsection, we evaluate it by exact values since we know all components of wavefunctions $\ket{{\rm S_i}}, \ket{{\rm S_j}}$ and the matrix elements of $R_\alpha$ by virtue of numerical simulations.

\begin{figure*}
    \centering
    \includegraphics[width=1.0\linewidth]{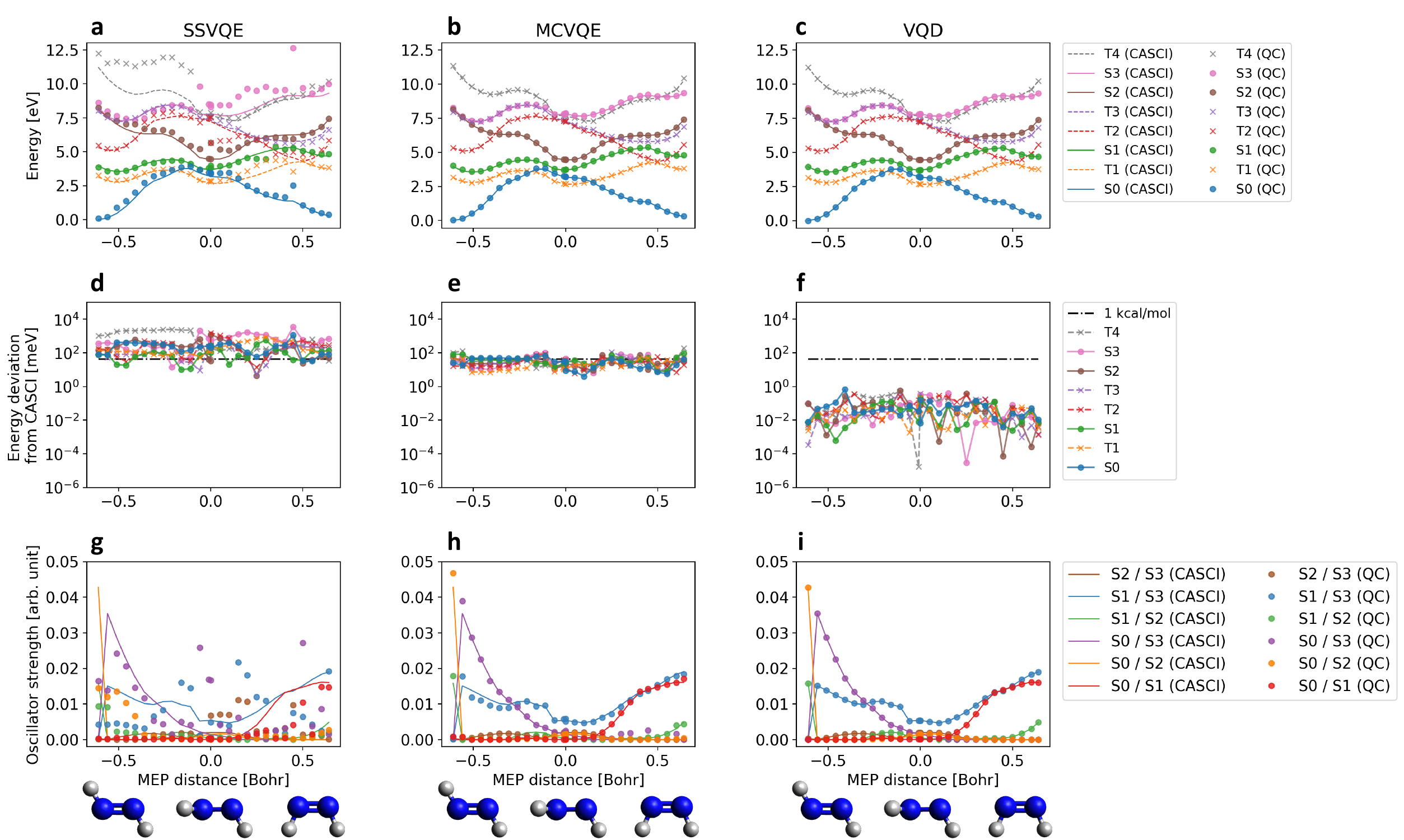}
    \caption{Potential energy curves along an S$_2$ MEP, from trans to cis structure of diazene calculated by (a) SSVQE, (b) MCVQE, and (c) VQD. Here, we set trans-S$_0$ minimum as zero energy, and the units of the MEP distance are mass-weighted coordinates divided with the square root of the total mass of the molecule.
    (d), (e), (f): The energy deviations of each method from the exact CASCI calculations. The dashed horizontal line denotes the ``chemical accuracy".
    (g), (h), (i): Oscillator strength between each pair of singlet states calculated by respective methods.
    Several atomic structures on the MEP are displayed at the bottom of the figure. }
    \label{fig:n2h2_compare}
\end{figure*}

Results for diazene along the MEP are shown in Fig. \ref{fig:n2h2_compare}.
Calculated energies and deviations from the exact ones (Fig. \ref{fig:n2h2_compare} (a-f)) show that the VQD method gives more accurate results than the other two in our simulation settings.
Moreover, from Fig. \ref{fig:n2h2_compare} (g, h, i), the VQD gives the most accurate oscillator strengths, which indicates that the method enables precise calculations of the wavefunctions as well as the energy spectrum.

\subsubsection{Azobenzene}
For AB, we perform VQE simulations using two structures: trans/cis isomers.
First, we obtain optimized trans/cis isomers using SA-CASSCF calculations, using an active space consisting of 3 orbitals ((lone pair on N) + $\pi$ (HOMO) + $\pi^*$ (LUMO)) with 4 electrons.
The S$_0$-S$_4$ and T$_1$-T$_3$ states are averaged in the SA-CASSCF calculations.
The structures used in the VQE simulations are the minimum energy structures of S$_0$ state.
Then, we perform the VQE simulations for the Hamiltonian in the same active space to obtain the eigenenergies and their deviations from the exact energies obtained by the exact diagonalization of the Hamiltonian. 
In this case, the number of qubits is six and we utilize RSP ansatz with $D=10$, where the total number of parameters is 50.
The oscillator strength is calculated as well by using the converged states. 
As the initial values of ansatz parameters, we use uniform random numbers within $[0, 2\pi]$.

\begin{figure*}
    \centering
    \includegraphics[width=1.0\linewidth]{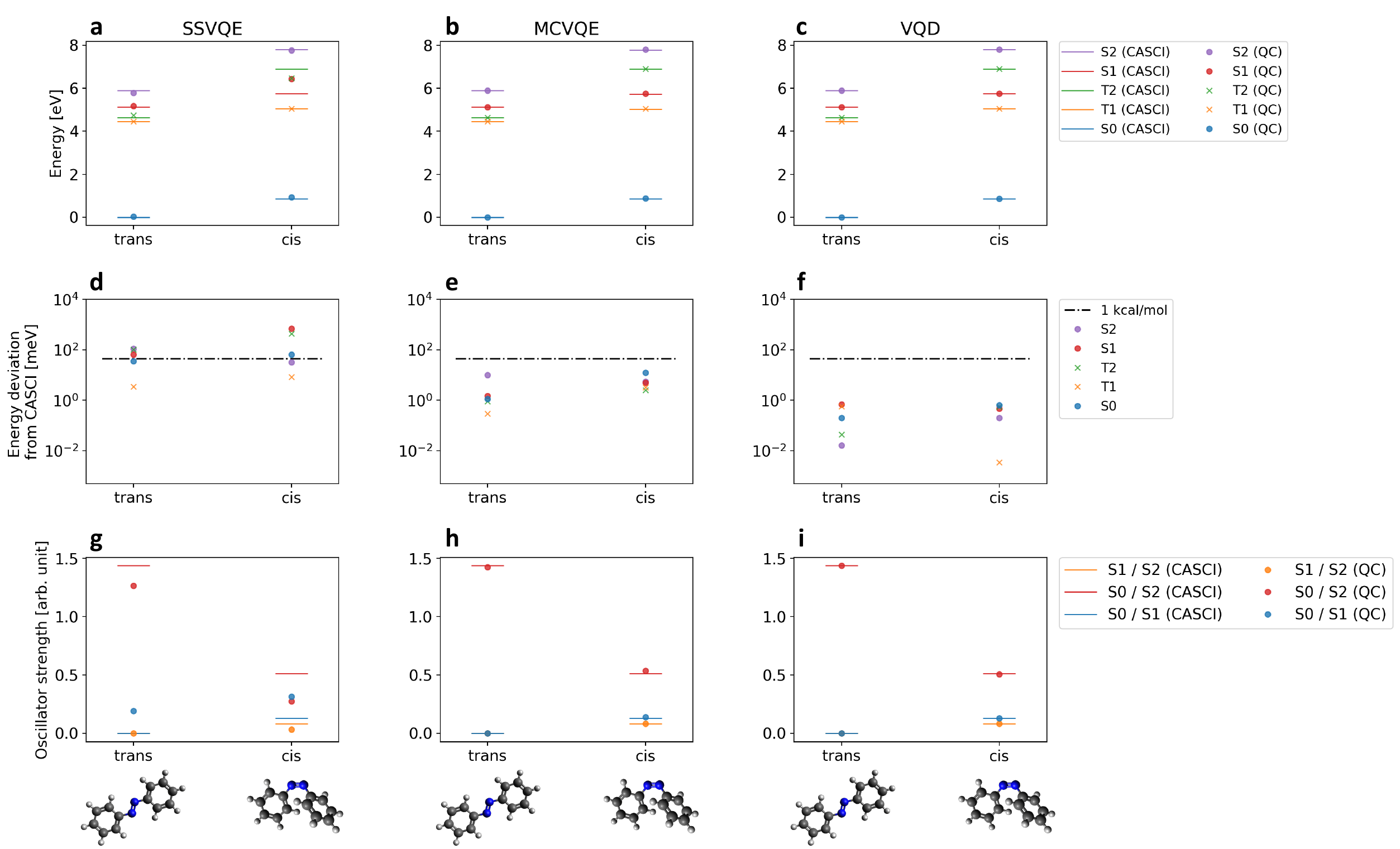}
    \caption{Energy levels of trans/cis AB calculated by (a) SSVQE, (b) MCVQE, and (c) VQD, where we set trans-S$_0$ state as zero energy.
    (d), (e), (f): The energy deviations of each method from the exact CASCI calculations. Dashed horizontal line denotes the ``chemical accuracy".
    (g), (h), (i): Oscillator strength between each pair of singlet states calculated by respective methods.
    The atomic structures are presented at the bottom of the figure. }
    \label{fig:azbz}
\end{figure*}

Results are shown in Fig. \ref{fig:azbz}.
Inheriting the trend from the previous results, the VQD  gives more accurate results than the other two.
Moreover, as evident in Fig.~\ref{fig:azbz} (g, h, i), the VQD gives the most accurate oscillator strengths, which indicates that the method enables us to generate precise wavefunctions.

\subsection{Discussion}
We have observed that the VQD has a better performance compared to the SSVQE and the MCVQE in this section. 
This result is probably because the requirement for the ansatz quantum circuit $U(\bm{\theta})$ is looser for the VQD than the SSVQE and the MCVQE; the ansatz with optimal parameters must make all reference states reside in the low-energy subspace {\it simultaneously} in the SSVQE and the MCVQE, while the VQD can do it {\it separately} for each reference state by the ansatz with different optimized parameters.

Moreover, the performance of the SSVQE seems worse than the MCVQE in our simulations.
The part of the reason is because we employ the ``weighted-sum" version of the SSVQE in the simulation~\cite{Nakanishi2019}.
If one used the ``equal-weight" version of the SSQVE, which is equivalent to the MCVQE, the difference will vanish as long as the optimization of the ansatz quantum circuit goes well.

\section{Implementation of transition amplitude evaluation for VQD \label{sec: demo method}}

The previous section demonstrates the significance of the technique presented in Sec. \ref{sec:eval}.
That is, we provide the hardware-friendly way to evaluate the transition amplitude between two eigenstates with the most accurate method for generating excited states among the previously proposed three methods.
In this section, to verify the correctness of the technique, we run it both on a more realistic simulator and the real hardware, which contain noise in expectation values of observables.

\begin{figure}
    \centering
    \includegraphics[width=\linewidth]{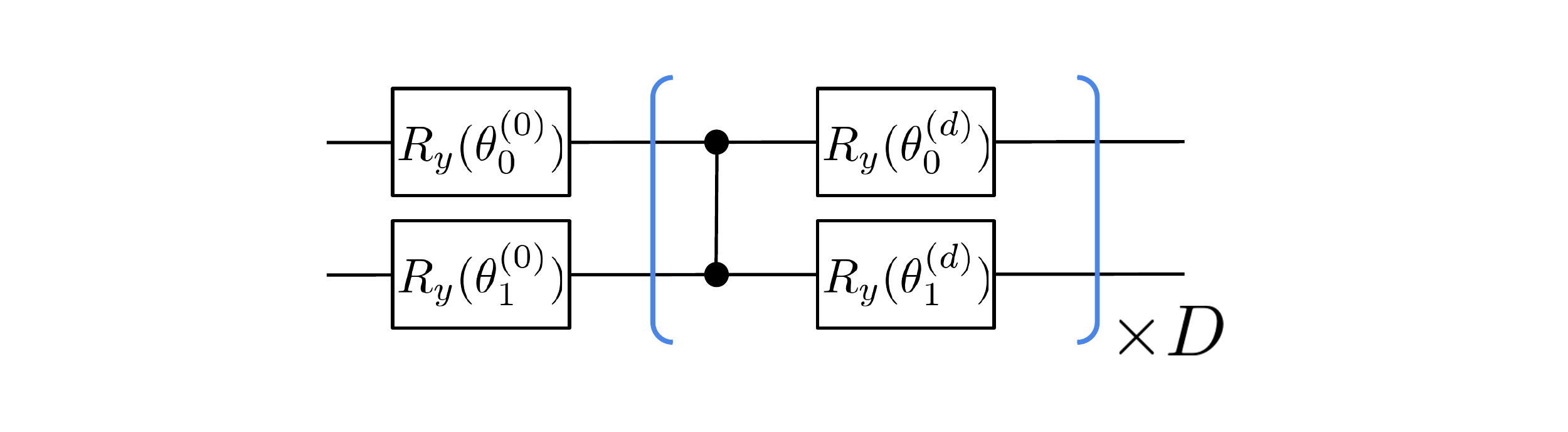} 
    \caption{``Two Local'' ansatz with $R_y$ rotation blocks and CZ entanglement blocks. The rotation angles in  $R_y(\theta)=\exp(-i\theta Y/2)$ are parameters to be optimized. $D$ denotes the depth of the circuit.}
    \label{fig:ry}
\end{figure}

We employ the molecular Hamiltonian of LiH and take the active space of (2e, 2o) with STO-3G minimal basis set.
We run the VQD simulation to calculate the potential energy curves of two singlet states, S$_0$ and S$_1$,
and calculate the oscillator strength between them by the method in Sec.~\ref{sec:eval}.
We take 36 points of the bond distances from 0.5 to 4.0 \AA. 
The parity-mapping \cite{Seeley2012} is used to map a fermionic (electron) Hamiltonian to a qubit Hamiltonian and to reduce the number of qubits from 4 to 2 utilizing the symmetry of particle number and $S_z$ \cite{bravyi2017tapering}.
As the ansatz for the VQD, we use the ``Two Local" with $R_y$ rotation blocks, CZ entanglement blocks, and a depth $D=4$, which is available in Qiskit Circuit library in Qiskit Terra v0.17.0 \cite{Qiskit} as \verb|qiskit.circuit.library.TwoLocal| (see Fig. \ref{fig:ry}).
For classical minimization of the cost function, the SLSQP method~\cite{Scipy2020} is employed for statevector or QASM simulations and the SPSA~\cite{spall1992multivariate} optimizer is used for real-hardware experiments.
Similar to the previous section, we set the convergence criterion so that the optimization terminates when the relative difference of energy expectation value between iterations becomes lower than $10^{-8}$.
Energy derivatives concerning the circuit parameters are calculated with the so-called parameter-shift rule~\cite{MitaraiQCL, Schuld2019} to mitigate the shot noise explained in the next paragraph.
To obtain only spin-singlet states, we add the expectation value of the total spin $\braket{S^2}$ as a penalty term to the cost function of the VQD.
The simulation in this section is performed using Qiskit \cite{Qiskit}.

\subsection{Sampling and Hardware simulations}
\begin{figure}
  \begin{minipage}[h]{1.0\linewidth}
    \centering
    \includegraphics[width=\linewidth]{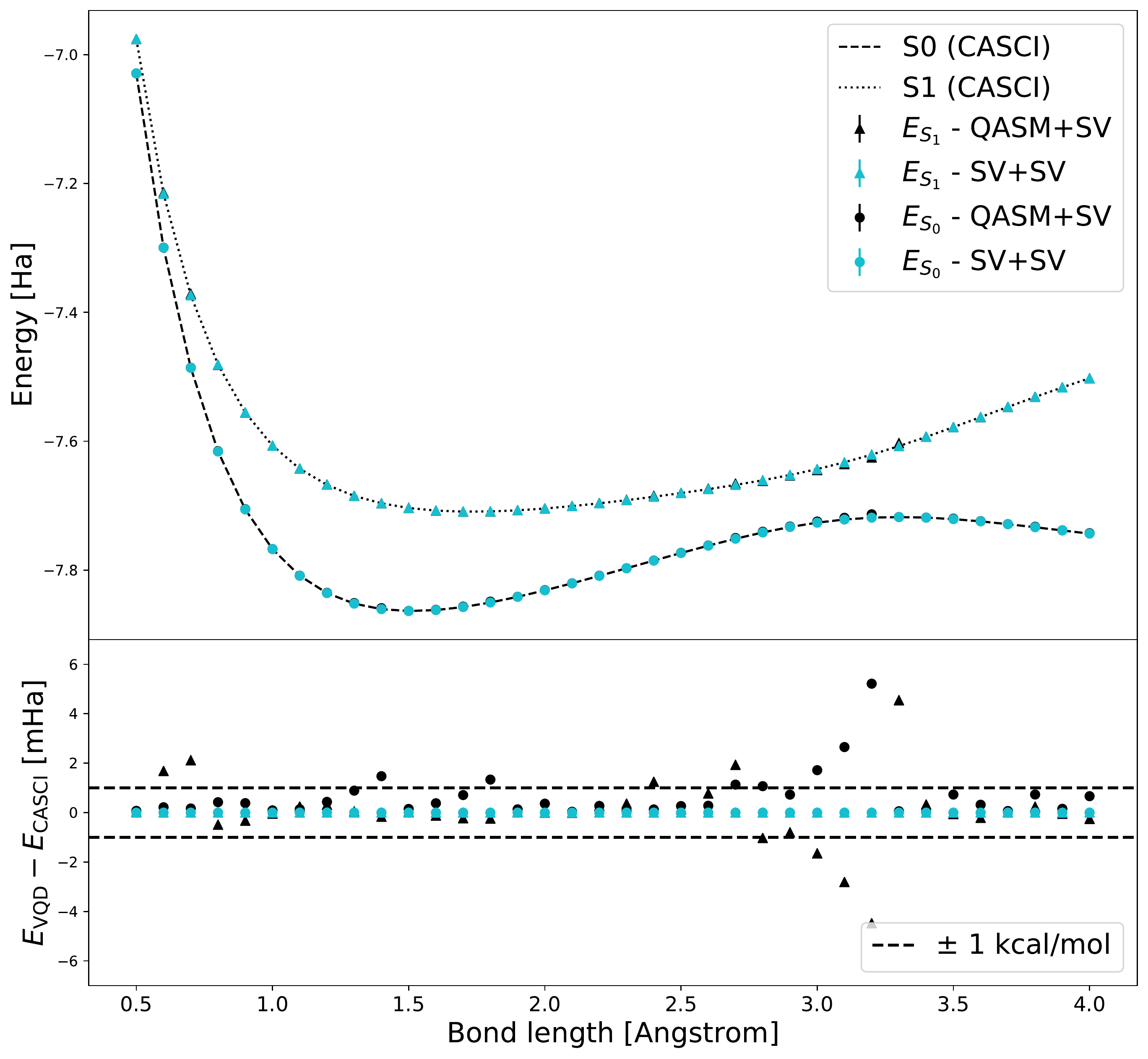}
  \end{minipage}\\
  \begin{minipage}[h]{1.0\linewidth}
    \centering
    \includegraphics[width=\linewidth]{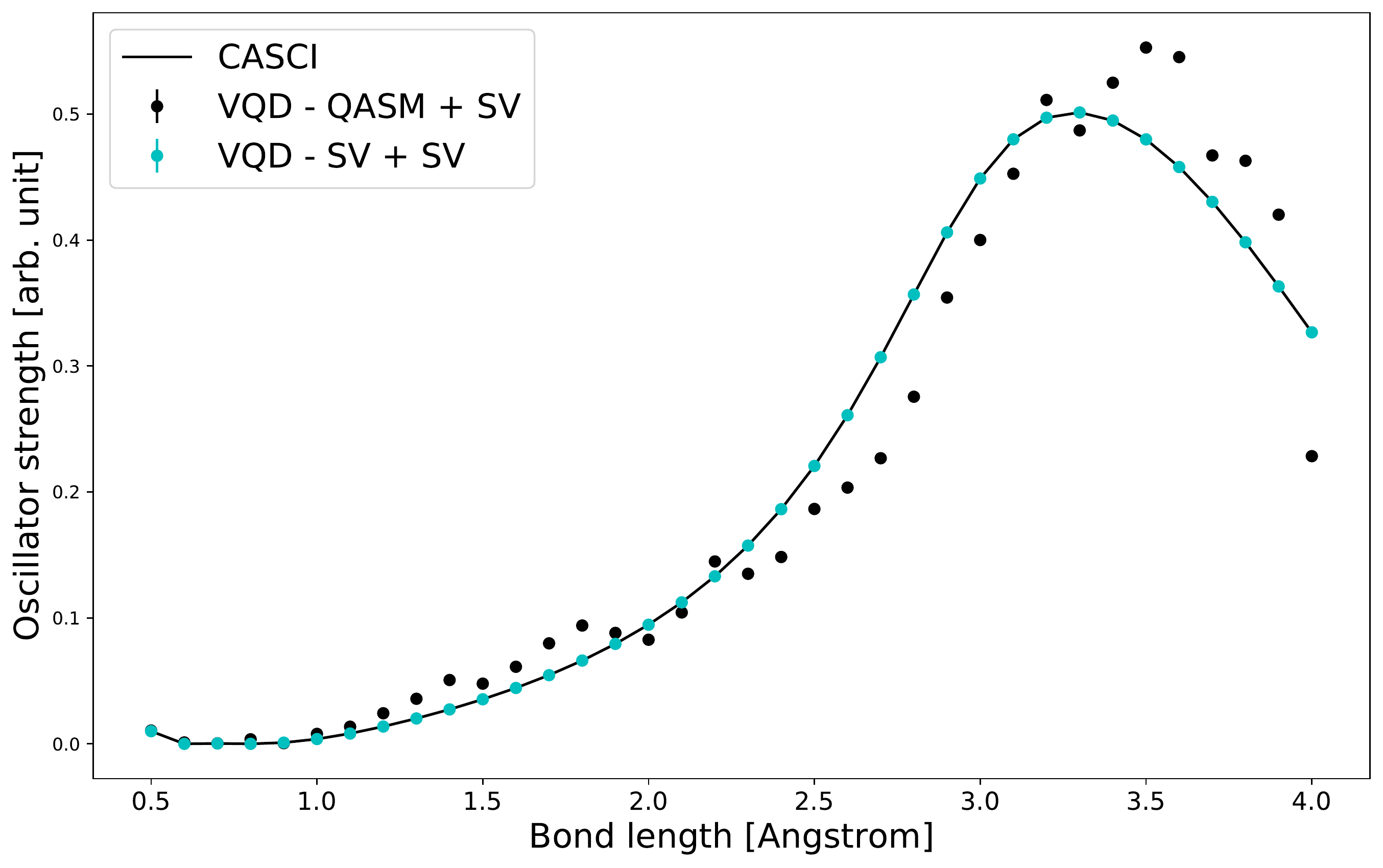}
  \end{minipage}
  \caption{Numerical simulation of LiH (2e, 2o) using the VQD. (top) Potential energy curve of S$_0$ and S$_1$ states and its deviation from exact CASCI calculations. (bottom) Oscillator strength between S$_0$ and S$_1$ states.
  For the VQD calculations, ``QASM+SV (SV+SV)" indicates that the numerical simulation optimizes the circuit parameters with (without) the shot+environmental noise. In both cases, we calculate oscillator strengths {\it without} the shot noise after the optimization. 
  For the VQD optimization routine in ``QASM+SV", we utilize the state purification technique described in the text.
  Error bars are calculated in a way explained in Appendix~\ref{appsec:errorbar}.}
  \label{fig:simulation_result}
\end{figure}

\begin{figure}
  \begin{minipage}[h]{1.0\linewidth}
    \centering
    \includegraphics[width=\linewidth]{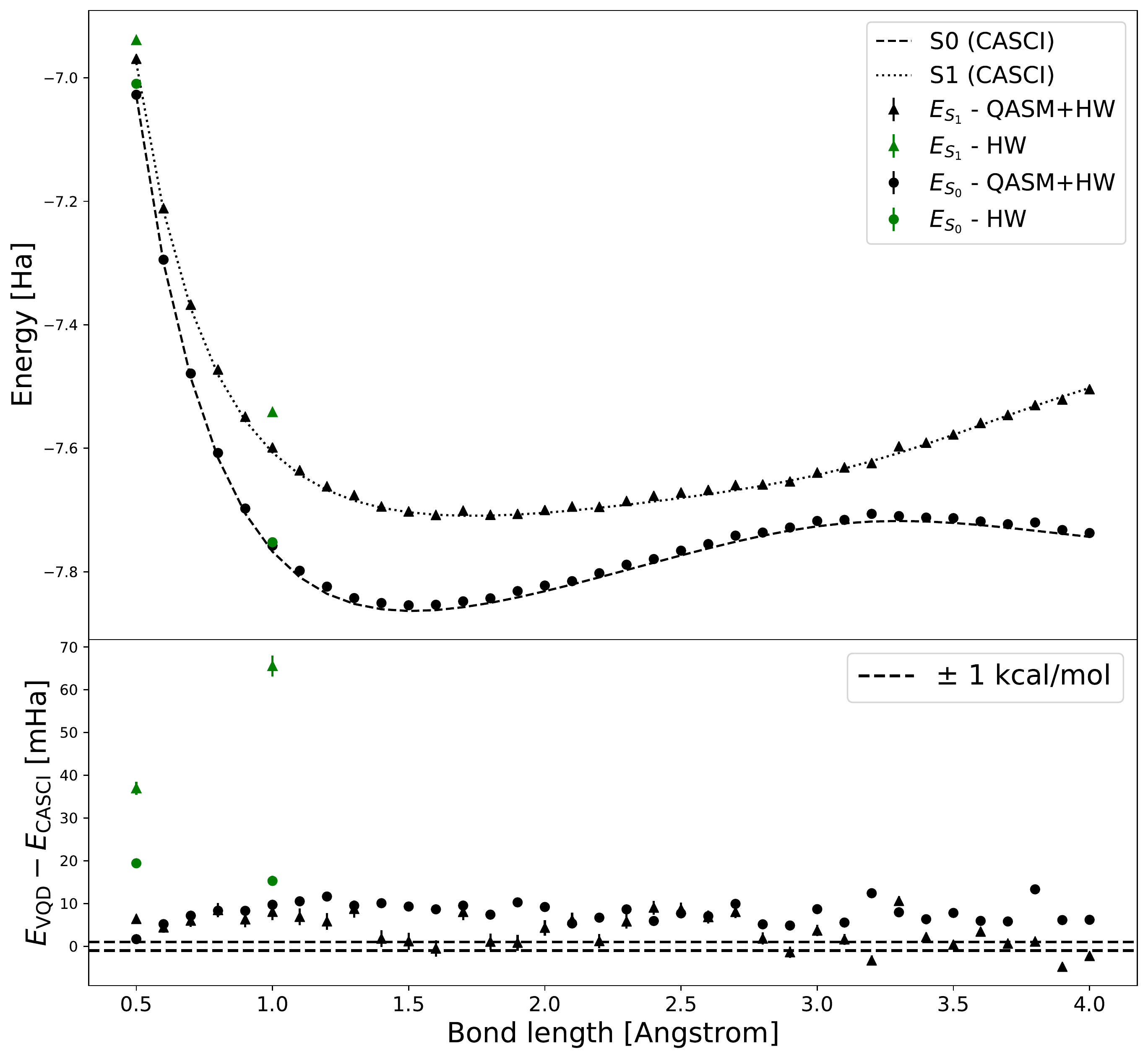}
  \end{minipage}\\
  \begin{minipage}[h]{1.0\linewidth}
    \centering
    \includegraphics[width=\linewidth]{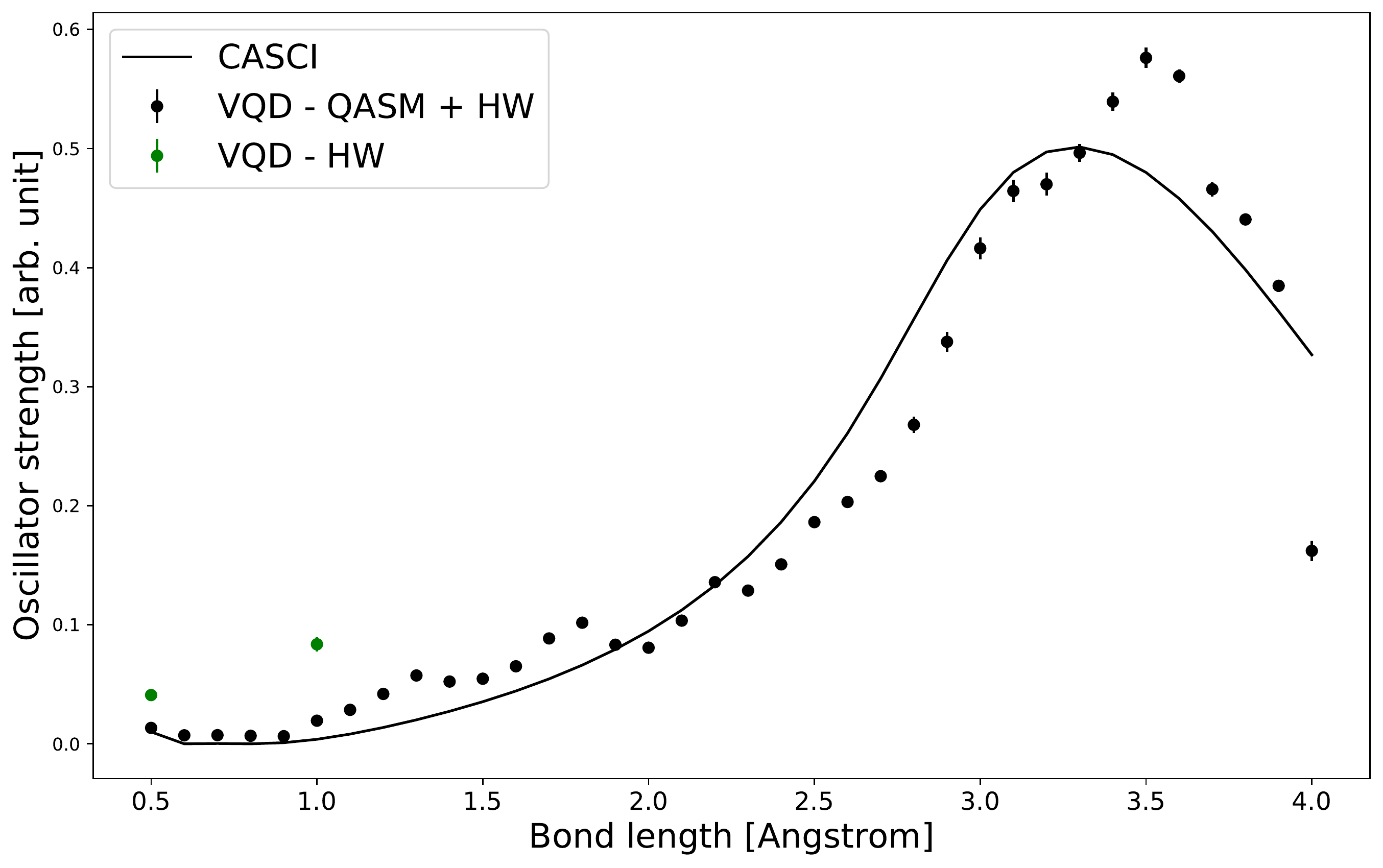}
  \end{minipage}
  \caption{Calculation result of LiH (2e, 2o) using the VQD on the real quantum device $ibmq \_ rome$. (top) Potential energy curve of S$_0$ and S$_1$ states and its deviation from exact CASCI calculations. (bottom) Oscillator strength between S$_0$ and S$_1$ states.
  Here, QASM+HW indicates that we first run the VQD optimization on a numerical simulator with shot+environmental noise, and then calculate the energy and the oscillator strength on the real device.
  On the other hand, simply ``HW" means that we run the whole process on the hardware.
  In both cases, we utilize the state purification technique described in the text for the VQD optimization routine.
  Error bars are calculated in a way explained in Appendix~\ref{appsec:errorbar}.}\label{fig:hardware_result}
\end{figure}

To elucidate where accuracy limitations arise in estimation of the transition amplitude, we evaluate simulating energy expectation values and overlaps in the optimization routine of the VQD in multiple approaches. First, we run the VQD optimization routine to obtain parameters using either an exact noiseless (statevector or SV) simulator or a backend noise model (QASM) simulator and then use these parameters in the statevector simulation (SV+SV or QASM+SV, respectively, in Fig.~\ref{fig:simulation_result}) to obtain physical quantities. Following this, we use the parameters obtained from simulation with a backend noise model and do the final evaluations on a quantum hardware $ibmq \_ rome$, or with the evaluations completely run on the $ibmq \_ rome$ (QASM+HW or HW, respectively, in Fig.~\ref{fig:hardware_result}). 

The statevector simulator is the same as explained in the previous section. For the noisy simulations, the expectation value of the energy is obtained by sampling 122,880 shots for each Pauli term in the Hamiltonian throughout the simulation. This sampling introduces the fluctuation in the energy expectation values, i.e., shot noise, as well as a realistic model of the quantum backend that includes T1,T2, gate and readout errors.
Similarly, 122,880 shots measurements are sampled for evaluating $|\braket{0|U_1^\dag U_2|0}|^2$ overlaps with our method in Sec.~\ref{sec:eval}. For hardware evaluations, we use the shot count of 8192.

In the VQD parameter optimization using either QASM simulator or the real quantum hardware, we utilized the so-called state purification technique. Due to the environmental noise, the quantum state generated by the ansatz circuit can be a mixed state instead of a pure state. This state can be purified by diagonalizing the density matrix of the state obtained from quantum state tomography technique and selection of the pure state with the maximum eigenvalue resulting from this process. The energy of the pure state is then refined by computing the expectation value of the Hamiltonian with respect to the obtained pure state. The values of parameters in ansatz corresponding to the purified state is determined by fitting the vector of the ansatz to the purified maximum eigenstate. For a thorough discussion on the purification approach, we refer the reader to the literature of Ref.~\cite{gao2021computational}.

In Fig.~\ref{fig:simulation_result}, we show the result of limitations in the parameter optimizations due to finite sampling and backend noise. We run experiments with the VQD iterative loop with the exact statevector simulations, as well as with a noise model to obtain the  ansatz parameters of Fig.~\ref{fig:ry}, and then evaluate the energies and oscillator strength using those parameters in a noise free statevector simulation as shown in Fig.~\ref{fig:simulation_result}.  When using the statevector simulator in the whole process (blue dots in Fig.~\ref{fig:simulation_result}), calculated energies achieve the exact solution. However, oscillator strengths deviate from the exact values when using the parameters obtained from the noisy simulations especially at bond lengths in which the energy spacing is small. This is, in part, because of the imperfection of the optimizations, as well as the imperfection of state purification, in which the ansatz generates a mixture of ground and excited state parameters. 

In Fig.~\ref{fig:hardware_result}, we show the result of the parameters obtained from noisy simulations with the final evaluation on cloud enabled quantum backend $ibmq \_ rome$, corresponding to the black dots for 36 different bond lengths. For two bond lengths, we carry out the simulation of energies and oscillator strength entirely on the quantum hardware, correspond to the green dots.  We reduce the impact of state preparation and measurement (SPAM) errors by preparing each of possible $2^2$ states and measuring the outcome to create an inversion matrix \cite{bravyi2020mitigating,barron2020measurement} after energy convergence is achieved (see Appendix~\ref{appsec:readout}). The initial parameters are selected at random when minimizing the ground state energy. These parameters are then used for the ground state parameters when minimizing the excited energy.   From Fig.~\ref{fig:hardware_result} we can see that the finite sampling on the real quantum hardware when determining the ansatz parameters (labeled HW in Fig.~\ref{fig:hardware_result}), leads to a more significant deviation from the exact answer when compared to the results that are obtained with a realistic backend noise model simulation, but with a significantly large number of shots (labeled QASM+HW in Fig.~\ref{fig:hardware_result}). 
This deviation of parameters on the real hardware can be relieved with the use of error mitigation techniques such as Richardson extrapolation \cite{kandala2019error,temme2017error}. 

\section{Conclusion}

In this work, we propose a general technique to evaluate transition amplitudes between two orthogonal states in a hardware-friendly manner on a quantum device.
Its immediate application is the evaluation of transition amplitudes between the approximate eigenstates of the Hamiltonian obtained by the VQD method.
The significance of the proposed method is supported by the comprehensive comparison of the three method, namely the SSVQE, the MCVQE and the VQD, in noiseless simulations which show the advantage of using the VQD for generating approximate excited states.
Finally, we also verify the proposed method by running it in on a real near-term quantum devices.
This work enlarges the possibility of the VQD and greatly advances the field of excited states calculations on a quantum device.
\section*{Acknowledgement}
This work was supported as part of a joint development agreement between Mitsubishi Chemical Corporation and QunaSys, and joint evaluation agreement between QunaSys and IBM.
A part of the numerical simulations in this work was done on Microsoft Azure Virtual Machines provided through the program Microsoft for Startups. A part of the quantum hardware simulations were done through the Startup provider as a part of the IBM Quantum Startup Network program.
YI, YON, TY, and KM, NE acknowledge Pauline Ollitrault, Wataru Mizukami and Tennin Yan for valuable discussions and Gilad Ben-Shach for a careful read of the manuscript.

\appendix
\section{Review of algorithms}\label{appsec:algorithms}

In this section, we provide a review of the algorithms used in the main text.

\subsection{VQE}

The VQE~\cite{Peruzzo2014} is a variational algorithm for finding the ground state of a system of $n$-qubits whose Hamiltonian is in the form of
\begin{equation}
    H = \sum_{P\in\{I,X,Y,Z\}^{\otimes n}} h_P P.
\end{equation}
where $I,X,Y,Z$ are single-qubit Pauli operators and $h_P \in \mathbb{R}$ is a coefficient.
If the number of terms with $h_P\neq0$ in the summation is not too large, we can evaluate the expectation value of the Hamiltonian, or energy of the system, by evaluating expectation values of each $P$ and then summing them on a classical computer.
To approximate the ground state of the Hamiltonian, the VQE uses parameterized quantum circuit $U(\bm{\theta})$, and iteratively optimize the parameter $\bm{\theta}$ so that the energy expectation value $E(\bm{\theta}):=\braket{0|U^\dagger(\bm{\theta})HU(\bm{\theta})|0}$, where $\ket{0}$ is an initialized state of the quantum computer, is minimized.
The VQE algorithm proceeds as follows:
\begin{enumerate}
    \item Define a quantum circuit $U(\bm{\theta})$ with parameters $\bm{\theta}$.
    \item Repeat the followings until the convergence of $E(\bm{\theta})$.
    \begin{enumerate}
        \item Generate a state $\ket{\psi (\bm{\theta} ) }:=U(\bm{\theta})\ket{0}$.
        \item Evaluate the energy $E(\bm{\theta})$ by measuring $\braket{0|U^\dagger(\bm{\theta})HU(\bm{\theta})|0}$.
        \item Update the parameter $\bm{\theta}$ to decrease $E(\bm{\theta})$.
    \end{enumerate}
\end{enumerate}
When the convergence is reached, we expect that $\ket{\psi(\bm{\theta})}$ and $E(\bm{\theta})$ is an approximate ground state and its energy from the variational principle of the quantum mechanics.

\subsection{Subspace-search VQE}\label{appsec:ssvqe}
The SSVQE \cite{Nakanishi2019} uses multiple initial states to search low-energy subspace of a Hamiltonian.
The SSVQE algorithm can be summarized as follows:
\begin{enumerate}
    \item Define an ansatz quantum circuit $U(\bm{\theta})$ and mutually orthogonal initial states (reference states) $\{\ket{\varphi_i}\}_{i=1}^k$. The reference states must be chosen so that one can readily make superpositions of them on a quantum device such as the computational basis.
    \item Repeat the following steps until the convergence.
    \begin{enumerate}
        \item Generate a set of states $\ket{\psi_i(\bm{\theta})}:=U(\bm{\theta})\ket{\varphi_i}$.
        \item Evaluate a cost function defined as a weighted sum of energies, $L_{\bm{w}}(\bm{\theta}):=\sum_{i=1}^k w_i \braket{\psi_i(\bm{\theta})|H|\psi_i(\bm{\theta})}$,
        where the weight vector $\bm{w}$ is chosen such that $w_1 > w_2 > \cdots > w_k > 0$. 
        \item Update parameter $\bm{\theta}$ to decrease $L$.
    \end{enumerate}
\end{enumerate}

The weight vector $\bm{w}$ has the effect of choosing which $\ket{\varphi_i}$ converges to which excited state.
The cost function $L_{\bm{w}}(\bm{\theta})$ reaches its global minimum when the ansatz circuit $U(\bm{\theta})$ maps $\ket{\varphi_i}$ to the $i$-th excited state $\ket{E_i}$ of the Hamiltonian.

We note here that the assumed ability to create the superposition of $\{\ket{\varphi_i}\}_i$ enables us to evaluate transition amplitudes of an operator between two eigenstates.
It can be performed by creating two different superpositions of two eigenstates, measuring the operator of the interest, and postprocessing on a classical computer \cite{Nakanishi2019}.

\subsection{Multistate contracted VQE}
The protocol of MCVQE \cite{Parrish2019} is similar to that of the SSVQE.
It works as follows:
\begin{enumerate}
    \item Perform Step 1 and 2 of the SSVQE, using a cost function where the weight vector is omitted:  $L(\bm{\theta}):=\sum_{i=1}^k  \braket{\psi_i(\bm{\theta})|H|\psi_i(\bm{\theta})}$.
    \item Using the converged $\bm{\theta}^*$, evaluate $\tilde{H}_{ij} := \braket{\psi_i(\bm{\theta})|H|\psi_j(\bm{\theta})}$ for all $i$ and $j$.
    \item Diagonalize the matrix $\tilde{H}=\{\tilde{H}_{ij}\}_{i,j=1}^k$ on a classical computer.
\end{enumerate}
Energy spectrum of $\tilde{H}$ approximates that of the original $H$, and approximate eigenstates are obtained by superposing $\{ \ket{\psi_i(\bm{\theta})} \}_i$ with coefficients determined by the eigenvectors of $\tilde{H}$. 
The evaluation of transition amplitudes between the approximate eigenstates can be performed in the same manner as the SSVQE.

\subsection{Variational quantum deflation}
The VQD algorithm \cite{Higgott2019} is probably the most straightforward way to construct approximate eigenstates of a Hamiltonian $H$.
The algorithm for finding the $k$-th excited state is as follows.
\begin{enumerate}
    \item Perform the VQE and obtain an optimal parameter $\bm{\theta}^*_0$ an approximate ground state $\ket{\psi(\bm{\theta}_0^*)}$.
    \item Set $j=1$ and repeat the following until $j=k$.
    \begin{enumerate}
        \item Define a Hamiltonian \begin{equation}
            H_j := H + \sum_{i=0}^{j-1} \beta_{i}\ket{\psi(\bm{\theta}_{i}^*)}\bra{\psi(\bm{\theta}_{i}^*)},
        \end{equation}
        where $\{\beta_i\}$ is a set of sufficiently large real-valued coefficient.
        \item Perform the VQE to find an approximate ground state of $H_j$.
        \item Increment $j$.
    \end{enumerate}
\end{enumerate}
The above algorithm works because $H_j$ has the $j$-th excited state of the original $H$.
To evaluate the expectation value of $H_j$, we need to evaluate the overlap between two states $\ket{\psi(\bm{\theta})}$ and $\ket{\psi(\bm{\theta}')}$.
It is suggested in \cite{Higgott2019} that we can either employ so-called destructive swap test \cite{Garcia2013} or measure them by $\left|\braket{0|U^\dagger(\bm{\theta})U(\bm{\theta}')|0}\right|^2$ exploiting the knowledge of the circuit.

We note that, while the previous two methods, namely the SSVQE and the MCVQE, can measure the transition amplitudes by creating the superposition of the initial states, there has been no efficient method for the VQD.

\section{Ancilla-based transition amplitude evaluation}\label{appsec:ancilla}
Here, we describe a method to evaluate $\braket{\psi_1|A|\psi_2}$ using an ancilla qubit.
We assume that we have descriptions of circuits $U_1$ and $U_2$ which generates $\ket{\psi_1}$ and $\ket{\psi_2}$ respectively from the initialized state $\ket{0}$.
Let,
\begin{align}
    \bar{\Lambda}(U_1) = \ket{0}\bra{0}\otimes U_1 + \ket{1}\bra{1}\otimes I, \\
    \Lambda(U_2) = \ket{0}\bra{0}\otimes I + \ket{1}\bra{1}\otimes U_2.
\end{align}
Then, we have the following equality,
\begin{align}
    &\textrm{Re}[\braket{\psi_1|P_i|\psi_2}] = \nonumber \\
    &(\bra{+}\otimes\bra{0})\bar{\Lambda}(U_1)^\dagger \Lambda(U_2)^\dagger (X\otimes P_i) \Lambda(U_2)\bar{\Lambda}(U_1) (\ket{+}\otimes\ket{0})\nonumber \\
    &\textrm{Im}[\braket{\psi_1|P_i|\psi_2}] = \nonumber \\
    &(\bra{+}\otimes\bra{0})\bar{\Lambda}(U_1)^\dagger \Lambda(U_2)^\dagger (Y\otimes P_i) \Lambda(U_2)\bar{\Lambda}(U_1) (\ket{+}\otimes\ket{0})\nonumber \\
\end{align}
We can recover $\braket{\psi_1|A|\psi_2}$ by combining them according to $\braket{\psi_1|A|\psi_2} = \sum_i a_i \braket{\psi_1|P_i|\psi_2}$.
However, the above method uses expensive controlled-$U$ gates, which might make it unfeasible on a near-term device.

\section{Statistical errors in shot noise simulator
\label{appsec:errorbar}}
The error bars of the energy of the S$_0$ and S$_1$ states in the top panel of Fig.~\ref{fig:simulation_result} and \ref{fig:hardware_result} are drawn by the outputs of \verb|evaluate_with_result| method of \verb|qiskit.aqua.operators.WeightedPauliOperator| of Qiskit Aqua v0.7.3~\cite{Qiskit}.
The output is calculated as follows.
For a Hamiltonian $H=\sum_{i=1}^N c_i P_i$ with $P_i$ being Pauli operator, its statistical sampling error is calculated by $\Delta H = \sqrt{\frac{1}{N} \sum_{i=1}^N c_i^2 (\Delta P_i)^2 }$,
where $(\Delta P_i)^2$ is the variance of measurement outcome of $P_i (=\pm1)$.

The error bars for the oscillator strength (bottom panel of Fig.~\ref{fig:simulation_result} and \ref{fig:hardware_result}), on the other hand, are calculated by propagation of the error for energies of S$_0$ and S$_1$ states described in the above and error in transition amplitude $ |\braket{{\rm S}_0 | R_\alpha | {\rm S}_1}|^2$, according to the definition of the oscillator strength (Eq.~\eqref{eq: def os}).
The latter error is estimated by five realized values of the transition amplitude obtained by using Eq.~\eqref{eq: main transition formula}, where each term in the right-hand side is computed as $|\braket{0|U|0}|^2$ (return probability of $\ket{0}$ state after some circuit $U$).

\section{Measurement error mitigation}
\label{appsec:readout}
State preparation and measurement errors can significantly impact the accuracy of hardware based results and, importantly, these error will not necessarily remain stable over the course of an experiment. To mitigate the impact of these errors we use the readout error mitigation technique described in \cite{bravyi2020mitigating}, in which $2^n$ state preparation circuits are run for n qubits. Specifically, we calibrate the $A_{\text{Full}}$:
\[\braket{y | A_{\text{Full}}| x} = \frac{m(y,x)}{N_{\text{cal}}} \]
where $x,y\in\{0,1 \}^n$ are the input and measured bitstring respectively, $m(x,y)$ is the number of shots for each input state $x$ and $N_{\text{cal}}$ is the total number of shots used. During the experiments on the real hardware, this matrix is calibrated every thirty minutes with 8192 shots by preparing $2^2$ basis states $\mathcal{C}=\{\{0,0 \},\{0,1 \},\{1,0 \},\{1,1 \} \}$. The matrix $A_{\text{Full}}$ is then inverted using the Qiskit function \verb|qiskit.ignis.mitigation.CompleteMeasFitter| and applied to the measured experimental outcomes to mitigate the impact of readout errors.

\section{Backend Properties}
\begin{table}[htbp!]
    \centering
    \caption{Summary of the properties of the backend, and simulation properties, including gate fidelities, qubits coherence times, and readout error for data presented in Sec. IV. $Fake Rome$ is the noise simulation of $ibmq \_ rome$ with the values shown in the table.}
    \label{tab:devices}
    \begin{tabular}{l r r r r r} \hline\hline
        & \multicolumn{3}{c}{Gate Errors ($\%$)} & & \\
        Qubits & CNOT & SX  &
         Readout & $\quad T_1$-times & $\quad T_2$-times \\
        &  &  & & $(\mu\mathrm{s})$ & $(\mu\mathrm{s})$  \\
        \\ \hline\hline
        $ibmq \_ rome$ \\
        q0, q1 & 0.732 & 0.027, 0.026 & 2.36, 3.67 & $86,\,94$ & $72,\,86$ \\
        $Fake Rome$ \\
        q0, q1& 0.781 & 0.0491, 0.0296 & 1.3, 3.0 & $86,\,103$ & $114,\,93$ \\
    \end{tabular}
\end{table}

\bibliographystyle{apsrev4-2}
\bibliography{90_library}

\end{document}